\newcommand{\rem}[1]{}
\newcommand{\beq}{\begin{equation}}
\newcommand{\eeq}{\end{equation}}
\newcommand{\beqa}{\begin{eqnarray}}
\newcommand{\eeqa}{\end{eqnarray}}
\newcommand{\ba}{\begin{array}}
\newcommand{\ea}{\end{array}}

\documentclass[pra,twocolumn,showpacs,preprintnumbers,
amsmath,amssymb,floatfix]{revtex4}
\usepackage{hyperref}
\usepackage[dvips]{epsfig}

\begin{document}

\title{Macroscopic Periodic Tunneling of Fermi Atoms
in the BCS-BEC Crossover}

\author{L. Salasnich$^{1,3}$, N. Manini$^2$ and F. Toigo$^3$}
\affiliation{$^1$CNR-INFM, Via Marzolo 8, 35131 Padova, Italy \\
$^2$Dipartimento di Fisica and CNISM,
Universit\`a di Milano, Via Celoria 16, 20133 Milano, Italy \\
$^3$Dipartimento di Fisica ``Galileo Galilei'' and CNISM,
Universit\`a di Padova, Via Marzolo 8, 35131 Padova, Italy}

\begin{abstract}
We study the macroscopic quantum tunneling of
two weakly-linked superfluids made of interacting fermionic atoms.
We derive atomic Josephson junction equations and
find that zero-mode and $\pi$-mode frequencies
of coherent atomic oscillations depend on the tunneling coefficient
and the sound velocity of the superfluid.
By considering a superfluid
of $^{40}$K atoms, we calculate these oscillation frequencies
in the crossover from the Bardeen-Cooper-Schrieffer state
of weakly-bound Cooper pairs to the Bose-Einstein Condensate
of strongly-bound molecular dimers.
\end{abstract}


\maketitle

\section{Introduction}

The crossover from the Bardeen-Cooper-Schrieffer (BCS)
state of Cooper fermion pairs
to the Bose-Einstein condensate (BEC)
of molecular dimers with ultra-cold two-hyperfine-components Fermi
vapors of $^{40}$K atoms \cite{greiner,regal,kinast}
and $^6$Li atoms \cite{zwierlein,chin,miller}
has been observed in the last years by several experimental
groups with the use of Fano-Feshbach resonances \cite{fano}.
From the theoretical side, extended BCS (EBCS) equations
\cite{eagles,leggett,Nozieres85}
have been used to reproduce density profiles \cite{perali} and collective
oscillations \cite{stringari,hu} of these Fermi gases.
In addition, more recent calculations
based on Monte Carlo (MC) fitting and superfluid
dynamics \cite{manini05} have shown that
the mean-field EBCS theory is quite accurate.

Recently, Spuntarelli {\it et al.}\ \cite{pieri-new} studied
the stationary Josephson effect \cite{josephson}
across the BCS-BEC crossover with neutral fermions by
using the EBCS equations:
in detail, they computed the current-phase relation
throughout the BCS-BEC crossover at zero temperature for
a two-spin component Fermi gas in the presence of a barrier.
The Josephson effect of atomic superfluids,
i.e.\ coherent oscillations between two weakly-linked
bosonic clouds, was predicted \cite{smerzi} and observed \cite{jo-exp}
with BECs. Josephson oscillations in superfluid atomic Fermi gases 
have been thoretically considered by Paraoanu {\it et al.} \cite{paraoanu}, 
Wouters {\it et al.} \cite{wouters}, and Adhikari \cite{adhikari}. 
The macroscopic oscillations 
of tunneling neutral atoms is closely related to the familiar
Josephson effect of charged electrons in superconductor junctions
\cite{barone} and its investigation can help a deeper understanding
of these exotic states of matter. 

In this work we present an investigation of
the dynamical Josephson effect in the BCS-BEC crossover based on a
time dependent local density approximation for the dynamics of
the Ginzburg-Landau (GL) order parameter of the atomic Cooper pairs
\cite{ginzburg}, starting from a reliable parameterization of the bulk
chemical potential \cite{manini05} in the crossover.
From our zero-temperature GL equation we obtain the
atomic Josephson junction equations for two weekly linked
fermionic superfluids. The main result provides zero-mode  
and $\pi$-mode Josephson frequencies of periodic quantum
tunneling as a function of the sound velocity.
Our investigation of the Josephson effect with neutral Fermi atoms, a
direct manifestation of macroscopic quantum phase coherence,
is of interest not only conceptually but also for future
applications in quantum computing \cite{averin,zagoskin}.

\section{Ginzburg-Landau equation at zero temperature}  

In a dilute Fermi gas of $N$ atoms with two equally populated spin
components and attractive inter-atomic strength at zero temperature,
superfluidity and coherence are strongly related to the properties of the
two-particle density matrix.
Let ${\hat \psi}_{\sigma}({\bf r},t)$ be the field operator that destroys a
fermion of spin $\sigma$ ($\sigma=\uparrow, \downarrow$) in the position
${\bf r}$ at time $t$.
The two-particle density matrix can be written as
\beqa
\langle
{\hat \psi}^+_{\sigma_1}({\bf r}_1,t)  
{\hat \psi}^+_{\sigma_2}({\bf r}_2,t)  
{\hat \psi}_{\sigma_1'}({\bf r}_2',t)  
{\hat \psi}_{\sigma_2'}({\bf r}_2',t)
\rangle \; =
\nonumber
\\
\sum_j N_j(t)
\chi_j^*({\bf r}_1 \sigma_1 , {\bf r}_2 \sigma_2, t)
\chi_j({\bf r}_1' \sigma_1' , {\bf r}_2' \sigma_2', t) \; ,
\eeqa
where $\langle \ \cdot \ \rangle$ is the ground-state average,
$\chi_j$ and $N_j$ are respectively normalized eigenfunctions and
eigenvalues of the two-particle density matrix, and the maximum
eigenvalue cannot exceed $N/2$ \cite{leggett,sala-odlro}.
If one eigenvalue, say $N_0$, is of order $N/2$, off-diagonal long-range
order is present.

At zero temperature, where the superfluid density coincides with the total
density, the GL order parameter describing
the motion of Cooper pairs of atoms is defined as
\beq
\Psi({\bf r},t) = \sqrt{N\over 2} \
\chi_0({\bf r}\!\uparrow , {\bf r}\!\downarrow , t)
= \sqrt{n({\bf r},t)\over 2} \, \exp{(i\theta({\bf r},t)}) \; ,
\eeq
where $n({\bf r},t)$ is the local atomic number density ($n({\bf r},t)/2$
is the local density of pairs) and $\theta({\bf r},t)$ is precisely the
phase of the condensate wave function $\chi_0$
\cite{ginzburg,leggett,landau}.
Note that $\theta({\bf r},t)$ is also the phase
of the gap function
$\Delta({\bf r},t)= |\Delta({\bf r},t)|
\exp{(i\theta({\bf r},t))}$ of
Cooper pairs \cite{ginzburg,leggett,sala-odlro,landau}.
Under an external potential $U({\bf r})$ acting on individual atoms,
the low-energy collective properties of the Fermi superfluid
\cite{manini05} can be described by the the following
highly nonlinear time-dependent GL equation (TDGLE) 
\beq
i \hbar {\partial \over \partial t} \Psi({\bf r},t)=
\left[ -{\hbar^2 \over 4m}\nabla^2 + 2 \ U({\bf r}) +  2 \ \mu(n({\bf r}))
\right] \Psi({\bf r},t)
\,.
\label{nlse}
\eeq
Here $m$ is the mass of one atom and
\beq
\mu(n)=\frac{\partial}{\partial n} (n\,{\cal E} (n))
\eeq
is the atomic 
bulk chemical potential of a homogeneous fluid with density $n$
\cite{landau2}, if ${\cal E}(n)$ is its energy per particle.
The phase of the order parameter drives the superfluid velocity
\beq
{\bf v}({\bf r},t) = {\hbar\over 2m}
\nabla \theta({\bf r},t) \; ,
\eeq
which is irrotational ($\nabla \wedge {\bf v}=0$)
by construction \cite{landau2}.
The equation of superfluid velocity permits to map Eq. (\ref{nlse}) to
the hydrodynamic equations of fermionic superfluids:
\beqa
{\partial n\over \partial t} &\!+\!& \nabla \cdot (n {\bf v}) = 0 \, ,
\label{sf-1}
\\
m {\partial {\bf v}\over \partial t} &\!+\!& \nabla
\left[ -{\hbar^2\over 8m}{\nabla^2 \sqrt{n}\over \sqrt{n} } +
{m\over 2}v^2 + U + \mu(n) \right] = 0 \,.
\label{sf-2}
\eeqa
In these zero-temperature hydrodynamical equations,
which are valid in the full BCS-BEC crossover to describe
macroscopic long-wavelength phenomena 
of the fermionic superfluid \cite{stringa-stro}, 
statistics enters through the equation of state $\mu(n)$ and 
the quantum-pressure term $-[\hbar^2/(8m \sqrt{n})]\nabla^2\sqrt{n}$,
which is absent in the classical hydrodynamic equations 
\cite{landau2,stringa-stro}. 
The coefficient $1/8$ in the quantum pressure holds for pairs of
fermions, as opposed to atomic bosons, where 
the coefficient is $1/2$, and ${\bf v}=(\hbar/m) \nabla \theta$. 
The bulk chemical potential $\mu(n)$ is the key
ingredient of our study; theoretical calculations 
in both asymptotic limits of $1/y$ and Montecarlo data \cite{astra} 
suggest that in the BCS-BEC crossover 
it can be written as
\beq
\mu(n) = {\hbar^2\over 2m}(3\pi^2\, n)^{2/3}
\left[ \epsilon(y) - {y\over 5}\epsilon'(y) \right]
,
\label{echem}
\eeq
where $\epsilon(y)$ is a dimensionless universal function
of the inverse interaction parameter
$y=1/k_F a_F$, where $k_F=(3\pi^2n)^{-1/3}$ is the Fermi wavenumber 
of the noninteracting fermions and $a_F$ is the fermion-fermion 
scattering length.
The function $\epsilon(y)$ was parameterized across the BCS-BEC crossover
by Manini and Salasnich \cite{manini05} to fit the MC data of
Astrakharchik {\it et al.}\ \cite{astra}
and the asymptotic expressions of the bulk energy per particle
\beq
{\cal E}={3\over 5} \epsilon_F \epsilon(y) \; ,
\eeq
with $\epsilon_F = (\hbar^2k_F^2/2m)$ the Fermi energy and
$k_F=(3\pi^2n)^{1/3}$ the Fermi wave number.
The parametrization of $\epsilon(y)$ chosen in Ref.~\cite{manini05}
is the following
\beq
\epsilon(y) = \alpha_1 - \alpha_2
\arctan{\left( \alpha_3 \; y \;
{\beta_1 + |y| \over \beta_2 + |y|} \right)}  \; ,
\eeq
where the values of the parameters
$\alpha_1,\alpha_2,\alpha_3,\beta_1,\beta_2$
are reported in that work. 
In the BEC regime ($y\gg 1$), where $\mu(n)\sim n$, the TDGLE reduces to the
time-dependent Gross-Pitaevskii equation for composite bosons of mass $2m$
subject to the effective potential $2 U({\bf r})+2\mu(n({\bf r},t))$
\cite{pieri-new}. In the BCS regime ($y \ll -1$), where 
$\mu(n)\sim \epsilon_F \left( 1 + k_Fa_F/(3\pi) \right)$, 
the TDGLE gives the Anderson-Bogoliubov 
mode (sound velocity) of neutral superconductors. 

Some years ago an effective nonlinear Schr\"odinger equation (NLSE) 
for superconductors was derived from the microscopic 
BCS Lagrangian in the low-frequency long-wavelength limit 
of $|\Delta({\bf r},t)|$ by Aitchison {\it et al.} 
and De Palo {\it et al.} \cite{ait}. That NLSE and our TDGLE 
produce the same Anderson-Bogoliubov mode \cite{ait,manini05}, 
but TDGLE describes accurately sound velocity and other 
collective modes of the fermionic superfluid 
in the full BCS-BEC crossover \cite{manini05}. 

\section{Dynamics of two weakly-linked Fermi superfluids}

Assume that $U({\bf r})$ is a potential with a barrier that splits the
superfluid into two subsystems A and B separated by a region C where the
modulus of the order parameter is exponentially small
\cite{notarella}.
We look for a time-dependent solution of the TDGLE of the form
\beq
\Psi({\bf r},t) = \Psi_{\rm A}(t) \ \Phi_{\rm A}({\bf r}) +
           \Psi_{\rm B}(t) \ \Phi_{\rm B}({\bf r}) \; ,
\eeq  
where $\Phi_{\alpha}({\bf r})$ is the quasi-stationary
solution (real and normalized to unity) of the TDGLE
localized in region $\alpha$ ($\alpha={\rm A},{\rm B}$).

By inserting this ansatz for $\Psi$
into Eq.~\eqref{nlse}, after integration over space
and neglecting exponentially small $\Phi_A \Phi_B$ terms,
the system can be described by the following
two-state model \cite{feynman}:

\beqa
i\hbar {\partial \over \partial t} \Psi_{\rm A} = E_{\rm A} \
\Psi_{\rm A} - K \ \Psi_{\rm B}
\label{tun-a}
\\
i\hbar {\partial \over \partial t} \Psi_{\rm B} = E_{\rm B} \
\Psi_{\rm B} - K \ \Psi_{\rm A}
\label{tun-b}
\eeqa
for the two complex coefficients $\Psi_{\alpha}(t)$.
Here $E_{\alpha} = E_{\alpha}^0+E_{\alpha}^I$
is the energy in region $\alpha$, with
\beq
E_{\alpha}^0 =
\int  \Phi_{\alpha}
\left[-{\hbar^2\over 4m}\nabla^2 + 2U
\right] \Phi_{\alpha} \, d^3{\bf r}
\eeq
and
\beq
E_{\alpha}^I \simeq  
\int  \Phi_{\alpha}
2\mu\left(2|\Psi_{\alpha}|^2\Phi_{\alpha}^2\right)\Phi_{\alpha}
\, d^3{\bf r} \; .
\eeq
The coupling energy $K$ describes phenomenologically
the tunneling between the two regions. It is quite common
to use a phenomenological tunneling energy
to analyze the Josephson effect in superconductors
\cite{leggett,barone,landau,feynman}.
In general, thermal \cite{caldeira}
and quantum \cite{zwerger} fluctuations will affect the value of $K$.

At zero-temperature the single-particle tunneling energy $t_0$
can be estimated as
\begin{equation}
t_0 \simeq -\frac 12
\int \Phi_{\rm A} \left[ -{\hbar^2\over 4m}\nabla^2 + 2U
\right]\Phi_{\rm B} \, d^3{\bf r}
\, ,
\end{equation}
if the barrier is orthogonal to the $x$ axis and such that
$\int_C \left[{8m\over \hbar^2}
\left(U(x)-\mu(n)\right)\right]^{1/2} dx \gg 1$.
In the BEC region the quantum depletion is negligible
\cite{sala-odlro}, the tunneling is fully coherent,  
and the coupling energy is simply $K \simeq 2t_0$
\cite{smerzi,zwerger}. Instead
in the deep BCS regime, where there is
a very large quantum depletion,
only a small fraction of pairs perform coherent
tunneling, and microscopic calculations suggest
$K \simeq |\Delta|t_0^2/(8\pi \epsilon_F^2 N)$ \cite{ambeg}.
Unfortunately a microscopic derivation of $K$
in the full BCS-BEC crossover is not yet available.

Under the assumption that the
double-well potential $U({\bf r})$ has the shape of two weakly-linked
sharp-edged boxes of volumes $V_{\rm A}$ and $V_{\rm B}$,
we can write $\Psi_{\alpha}(t) = \sqrt{N_{\alpha}(t)/2}
\exp{\left(i\theta_{\alpha}(t)\right)}$,
where $N_{\alpha}(t)$ and $\theta_{\alpha}(t)$ are
the number of fermions and the phase in region $\alpha$.
In terms of the phase difference
\beq
\varphi(t) =\theta_{\rm B}(t)-\theta_{\rm A}(t)
\eeq
and relative number imbalance
\beq
z(t) ={N_{\rm A}(t)-N_{\rm B}(t) \over N} \; ,
\eeq
with $N=N_{\rm A}(t) + N_{\rm B}(t)$ the constant total number of atoms,
Eqs.~\eqref{tun-a} and \eqref{tun-b} give
\beqa
{\dot z} &=& - {2K\over \hbar} \, \sqrt{1 - z^2} \, \sin\varphi
\, ,
\label{ajj-1}
\\
{\dot \varphi} &=&
{2\over \hbar}\left[
\mu\left({N\over 2V_{\rm A}}(1+z)\right) -
\mu\left({N\over 2V_{\rm B}}(1-z)\right)
\right]
\nonumber
\\
&+& \! {2K\over \hbar} \, {z \over
\sqrt{1 - z^2}} \, \cos\varphi +
{E_{\rm A}^0 - E_{\rm B}^0 \over \hbar}
\, .
\label{ajj-2}
\eeqa
These are the atomic Josephson junction (AJJ) equations  
for the two dynamical variables
$z(t)$ and $\varphi(t)$ describing the oscillations
of $N$ Fermi atoms tunneling in the superfluid state between
region $A$ of volume $V_{\rm A}$ and region $B$ of volume $V_{\rm B}$.
These equations, linking the tunneling current
$
I = - {\dot z} N/2 =
(KN/\hbar)\,\sqrt{1 - z^2}\,\sin\varphi =
I_0 \, \sqrt{1-z^2} \, \sin\varphi  
$
to the phase difference $\varphi$,
reduce to the familiar Josephson's expression
$I=I_0\sin(\varphi)$, well established for BCS superconductors,
in the appropriate limit $|z|\ll 1$ \cite{barone}.
Also, in the deep BEC regime, where $\mu(n)\sim n$,
the AJJ equations reduce to the bosonic
Josephson junction (BJJ) equations introduced
by Smerzi {\it et al.}\ \cite{smerzi} and found to describe
accurately the density oscillations of weakly linked condensates
\cite{jo-exp}.

\section{AJJ equations: linear regime}

The AJJ equations can be solved across the BCS-BEC crossover,
even taking into account a small but finite imbalance $z$.
To first order in $z$, Eqs. (\ref{ajj-1}) and (\ref{ajj-2}) read
\beqa
{\dot z} &=& - {2 K\over \hbar} \sin{\varphi}
\,,
\label{eq-a}
\\
{\dot \varphi} &=& {2 \Lambda \over \hbar} z +
{2K\over \hbar} z \cos{\varphi} + {2 {\tilde \mu}_{\rm AB}\over \hbar}
\, ,
\label{eq-b}
\eeqa
where
\beq
\Lambda = {N\over 2} \left[{1\over V_{\rm A}}
{\partial \mu\over \partial n}\Big|_{\rm A}
+ {1\over V_{\rm B}} {\partial \mu\over \partial n}\Big|_{\rm B} \right]
,
\label{lambda}
\eeq
and
\beq
{\tilde \mu}_{\rm AB} = {1\over 2} (\mu|_{\rm A} - \mu|_{\rm B})
+ (E_{\rm A}^0 - E_{\rm B}^0) \; ,
\eeq
with $\mu|_{\alpha}$ and
${\partial \mu\over \partial n}\Big|_{\alpha}$
the bulk chemical potential and its derivative are
calculated at the density $N/(2V_{\alpha})$, $\alpha=A,B$.

In the symmetric case
($V_{\rm A}=V_{\rm B}=V/2$ and $E_A^0=E_B^0$)
$\Lambda$ takes the particularly simple form
\beq
{\Lambda} = 2 n \,{\partial \mu \over \partial n}
=  2 m c_s^2 \; ,  
\eeq
where $c_s$ is the sound velocity
computed at the mean density $n=N/V$ of the superfluid.
>From Eq. (\ref{echem}) one has thus
\beq
\Lambda
= 2 m v_F^2
\left[ {1\over 3} \epsilon(y) - {y\over 5}\epsilon'(y) +
{y^2\over 30} \epsilon''(y) \right]
,
\label{sound}
\eeq
where $v_F=\hbar k_F/m$ is the Fermi velocity
of non-interacting fermions.

In the symmetric case ${\tilde \mu}_{AB}=0$,
Eqs.~(\ref{eq-a}) and (\ref{eq-b}) admit a stable
stationary solution $({\bar z},{\bar \varphi})$
with ${\bar z}=0$ and ${\bar \varphi}=2\pi j$, for integer $j$.
If $\Lambda/K<1$, also ${\bar z}=0$ and ${\bar \varphi} = \pi (2j+1)$
is a stable stationary solution.
These stationary solutions remain stable even if one includes higher-order
terms in the $z$-expansion of Eqs.~\eqref{ajj-1} and \eqref{ajj-2}.
Small oscillations around the stable
solutions with ${\bar \varphi} = 2 \pi j$ and ${\bar \varphi}
= \pi (2j+1)$ have frequencies
\beq
\nu_{0/\pi} = \frac {K}{\pi \hbar} \sqrt{1 \pm {\Lambda\over K}}
\,,
\label{figata}
\eeq
and are called zero-mode (with $+$) and $\pi$-mode (with $-$),
respectively.
The zero-mode is the analog
of the Josephson plasma frequency
in superconducting junctions \cite{barone}.  
The analogous of this $\pi$-mode was observed
in weakly coupled reservoirs of
$^3$He-B \cite{backhaus} and discussed
in the BJJ equations \cite{smerzi}.

\begin{figure}
{\includegraphics[width=7. cm,clip]{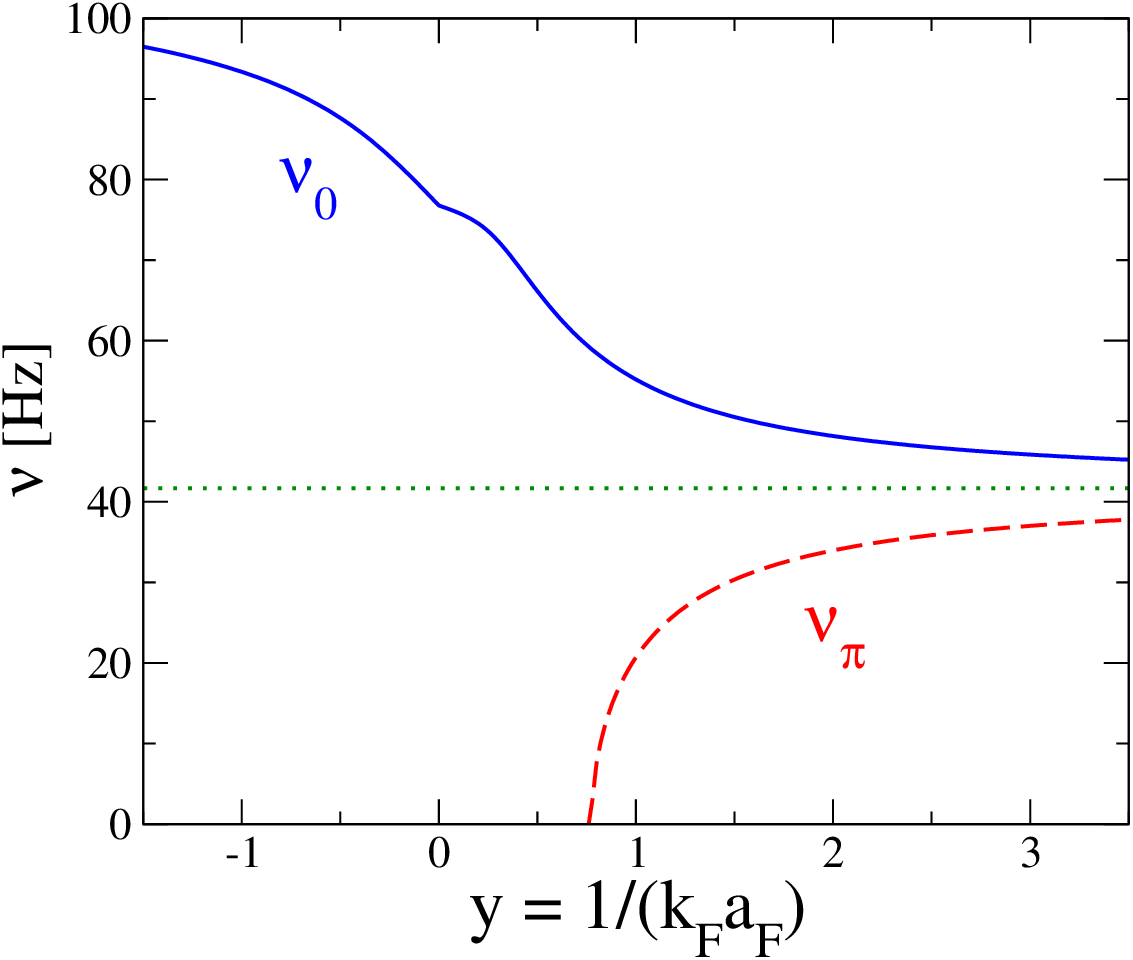}}
\caption{(color online).
Zero-mode frequency $\nu_0$  (solid) and
$\pi$-mode frequency $\nu_{\pi}$ (dashed)
for a superfluid of $N=10^6$ $^{40}$K atoms
between two symmetric regions of volume $25\cdot 10^6~\mu$m$^3$ (mean
density $n=0.02$~$\mu$m$^{-3}$) and fixed tunneling parameter
$K/k_{\rm B} = 10^{-9}$\, Kelvin.}
\label{fig-luca}
\end{figure}
By using the functional dependence on the inverse interaction parameter
$y=1/(k_Fa_F)$ of the sound velocity $c_s$, and inserting this expression
of $\Lambda$ in Eq.~(\ref{figata}), we obtain the oscillation frequencies
$\nu_0$ and $\nu_{\pi}$.
As an example, consider a fermionic superfluid of $^{40}$K atoms with
total density $n=0.02$ atoms/$\mu$m$^3$: Fig.~\ref{fig-luca} reports
$\nu_{0/\pi}$ as a function of $y$, assuming that the tunneling energy
can be held fixed to the value $K/k_{\rm B} =10^{-9}$ Kelvin
(where $k_{\rm B}$ is the Boltzmann constant).
As previously stressed in the deep BCS regime $K$ is proportional
to $|\Delta|$ and becomes rapidly very small, thus in practice
it would be difficult to keep constant.  
The zero-mode frequency decreases for increasing $y$ across the crossover,
approaching the asymptotic value $K/(\pi\hbar)$ (dotted line) in the BEC
limit ($y\gg 1$).
In contrast, the $\pi$-mode does not exist until past the unitarity point
$y=0$, i.e.\ until the value of $\Lambda$ becomes smaller than $K$; then
$\nu_\pi$ increases approaching the same large-$y$ limit as $\nu_0$.

In limiting cases of the BCS-BEC crossover analytical expressions
are available for $\Lambda$.
(i) {\it Deep BCS regime} ($y\ll -1$):
the sound velocity $c_s$ approaches $v_F/\sqrt{3}$ \cite{manini05},
and $\Lambda = 2 m v_F^2 = 2 (\hbar^2/m) (3\pi^2\, n)^{2/3}$.
(ii) {\it Unitarity point} ($y=0$):
according to MC simulations of Astrakharchik {\it et al.}~\cite{astra}
the sound velocity is $c_s \simeq 0.37 \ v_F$, thus
$\Lambda = 0.29\, m v_F^2$.
(iii) {\it Deep BEC regime}:
all molecular dimers are in the BEC and the sound velocity is such
that $mc_s^2 = \pi\hbar^2 a_M n/m$ \cite{manini05}, where $a_M=0.6\ a_F$
is the dimer-dimer scattering length \cite{astra}.
In this limit $a_F\to 0^+$ for non-interacting bosonic dimers,
$\Lambda = 0$ and corresponding $\nu_0 = \nu_{\pi}=K/(\pi\hbar)$.

\section{AJJ equations: nonlinear effects}

We come now to nonlinear effects in the AJJ equations (\ref{ajj-1}) and
(\ref{ajj-2}) in the symmetric case ($V_A=V_B=V/2$ and $E_A^0=E_B^0$).
The variables $\varphi$ and $z$ are canonically conjugate, with  
\beqa
{\dot z}&=&-{\partial H\over \partial \varphi} \; ,
\\
{\dot \varphi}&=&{\partial H\over \partial z} \; ,
\eeqa
where the Hamiltonian is
\beq
H={2\over \hbar} G(z)- {2K\over \hbar} \sqrt{1-z^2}\cos\varphi \; ,
\eeq
with
\beq
G'(z) = F(z)=\mu\!\left({N\over V}(1+z)\right) -
\mu\!\left({N\over V}(1-z)\right) \; .
\eeq
It is straightforward to show that these nonlinear equations
admit a symmetry-breaking
stationary solution $({\bar z},{\bar \varphi})$, with
${\bar \varphi}=\pi(2j+1)$
and with ${\bar z}$ the
solution of the equation $F(z)=K\,z/\sqrt{1-z^2}$.
The analysis of stability shows that this symmetry-breaking
solution, where the superfluid displays macroscopic
self-trapping, is stable only if $K>F'(\bar z)(1-\bar z^2)^{3/2}$.

\begin{figure}
{\includegraphics[width=8.2cm,clip]{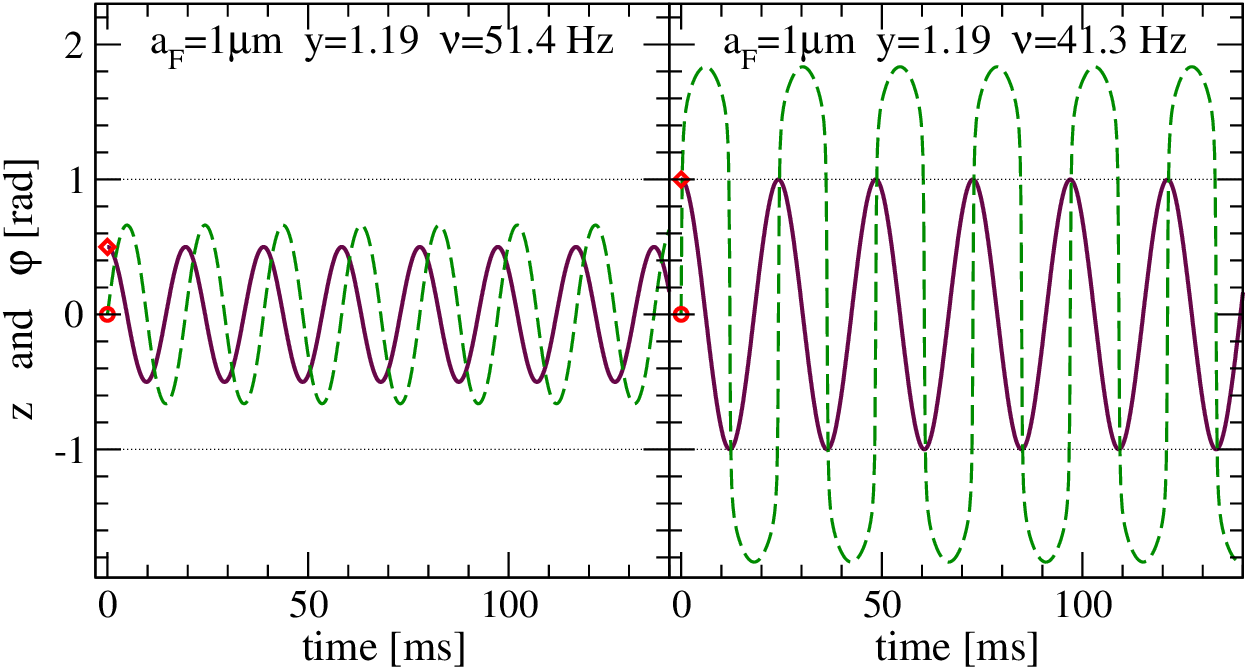}}
\caption{(color online).  
Nonlinear effects on the zero-mode oscillation of $N=10^6$ $^{40}$K atoms
between two symmetric regions with the same conditions of
Fig.~\ref{fig-luca} at fixed interaction $a_F=1$~$\mu$m,
corresponding to $y=1.19$.
These trajectories result from the integration of the coupled equations
\eqref{ajj-1} and \eqref{ajj-2} for $z(t)$ (solid) and $\varphi(t)$
(dashed), with initial conditions $\varphi(0)=0$, and $z(0)=0.5$ (left) or
$z(0)=0.999$ (right). }
\label{josephson_oscillations}
\end{figure}

\begin{figure}
{\includegraphics[width=8.2cm,clip]{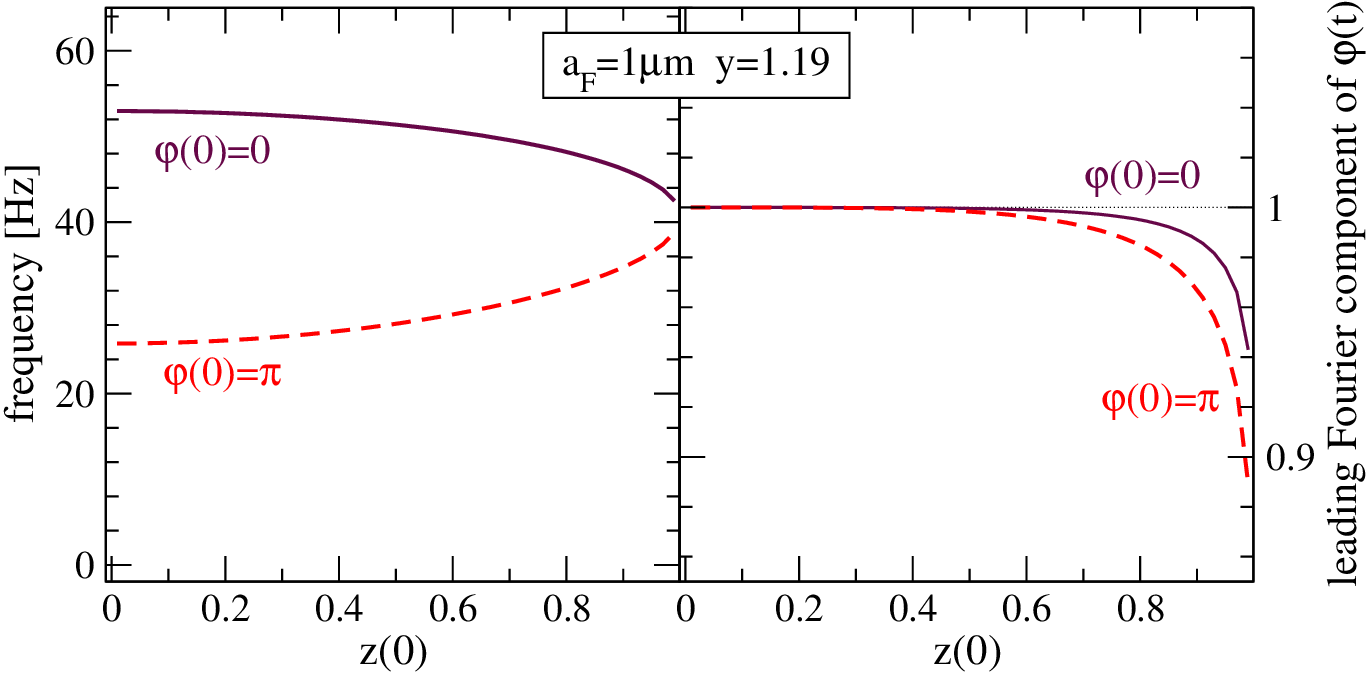}}
\caption{(color online).
Nonlinear effects on the zero-mode (solid) and $\pi$-mode (dashed)
frequency (left) and waveform (right) for the same conditions as in
Fig.~\ref{josephson_oscillations}, as functions of the initial amplitude
$z(0)$.
In an expansion
$\varphi(t) = \varphi(0)+\sum_{k=1}^{\infty} a_k \sin(2 \pi \nu k t)$,
the fractional leading Fourier component of the right panel is
$a_1^2/\sum_k a_k^2$.}
\label{josephson_freqANDfourier}
\end{figure}

To analyze nonlinear effects we solve numerically the AJJ equations.
Regular zero-mode oscillations of $z(t)$ and $\varphi(t)$ are displayed in
Fig.~\ref{josephson_oscillations}, under the same conditions as for
Fig.~\ref{fig-luca}.
The oscillation starting from $z(0)=0.5$ indicates that the solution
\eqref{figata} of the linearized equations \eqref{eq-a} and \eqref{eq-b}
are fairly accurate even for finite and not quite small amplitude.
Eventually however, for very large amplitude, $z(0)=0.999$, deviations
from the harmonic approximation become quite visible.
These deviations are illustrated, for increasing amplitude, in
Fig.~\ref{josephson_freqANDfourier}: the left panel shows that the
frequencies of the zero- and $\pi$-modes approach each other.
The right panel shows the decay in the fractional leading Fourier component
of the oscillation $\varphi(t)$, as the wave shape distorts from a perfect
sinusoid, and acquires higher (odd) Fourier components.
Note the extremely small deformations from perfect harmonic oscillations
and the minor deviations of the frequencies from the linear values
$\nu_0=53.0$~Hz and $\nu_\pi=25.8$~Hz of Eq.~\eqref{figata}, until
$z(0)\simeq 0.6$.
Similar evolutions near the $y=0$ point on the BEC side, and on the BCS
side, show oscillations of $z(t)$ around the self-trapping
symmetry-breaking stationary
solution $\bar z$, accompanied by monotonously increasing
phase $\varphi(t)$.

\section{Conclusions}

We have investigated the macroscopic quantum tunneling of
two weakly-linked Fermi superfluids in the BCS-BEC crossover
by using AJJ equations.
We remark that the coupling energy $K$ appearing
in our Josephson equations is a phenomenological
parameter: from the experimental measurement of the frequencies
of periodic quantum tuneling one can 
infer the value of $K$ by using Eq.\ (\ref{figata}).
Analytical expression of $K$ based on microscopic theory are available only
in the deep BCS regime and in the deep BEC regime.
An important issue is surely the development of a
microscopic theory of tunneling in the full BEC-BEC crossover.
We also stress that all Josephson oscillatory frequencies discussed
here cannot exceed the frequency $|\Delta|/(\pi\hbar)$, associated to the
breaking of Cooper pairs. 
The AJJ equations can be extended to investigate
Josephson junction arrays for neutral fermionic atoms in optical lattices,
potentially of extreme relevance for quantum information and quantum
computing with ultracold atoms.

\section*{Acknowledgments}

This work has been partially supported by Fondazione CARIPARO
and by the EU's 6th Framework Programme through
contract NMP4-CT-2004-500198.
L.S. has been partially supported by GNFM-INdAM and thanks
A. Bulgac, B.A. Malomed, A. Minguzzi, A. Parola, P. Pieri,
A. Smerzi and A. Trombettoni for useful discussions.


\begin{thebibliography}{99}
\bibitem{notarella} Notice that Spuntarelli, Pieri and 
Strinati \cite{pieri-new} 
do not work in the regime of weak-link: in their calculations
the energy barrier is much smaller than the Fermi energy.
\end{thebibliography}
\end{document}